%% file: elsarticle-template-num.tex
\documentclass[final,12pt]{elsarticle}

\usepackage{amsmath,amssymb,amsfonts}
\usepackage{algorithmic}
\usepackage{graphicx}
\usepackage{textcomp}
\usepackage{subcaption}
\usepackage{tabularx}
\usepackage{xcolor}
\usepackage{xspace}
\usepackage[inline]{enumitem} 
\usepackage{booktabs}
\usepackage{multirow}
\usepackage{graphicx}
\usepackage{hyperref}
\graphicspath{{img/}}
\usepackage{pifont}
\usepackage{soul}
\usepackage{booktabs}
\usepackage{lscape} 
\usepackage{array}
\usepackage{rotating}
\usepackage{color, colortbl}
\definecolor{LightCyan}{rgb}{0.88,1,1}
\definecolor{Red}{rgb}{1,0,0}

\newcommand*\rot{\rotatebox{90}}
\newcommand*\OK{\ding{51}}

\newcommand{\revision}[1]{#1\xspace}
\newcolumntype{H}{>{\setbox0=\hbox\bgroup}c<{\egroup}@{}}
\newcommand{\dscat}[1]{`#1'\xspace}

\journal{Journal of Systems and Software}

\begin{document}

\begin{frontmatter}



\title{Antipatterns in Software Classification Taxonomies}


\author[inst1]{Cezar Sas\corref{cor1}}
\ead{c.a.sas@rug.nl}
\cortext[cor1]{Corresponding Author}
\author[inst1]{Andrea Capiluppi}
\ead{a.capiluppi@rug.nl}

\affiliation[inst1]{organization={Bernulli Institute, University of Groningen},
            addressline={Nijenborgh 9}, 
            city={Groningen},
            postcode={9747 AG}, 
            country={Netherlands}}

\begin{abstract}

Empirical results in software engineering have long started to show that findings are unlikely to be applicable to all software systems, or any domain: results need to be evaluated in specified contexts, and limited to the type of systems that they were extracted from. This is a known issue, and requires the establishment of a classification of software types.

This paper makes two contributions: the first is to evaluate the quality of the current software classifications landscape. The second is to perform a case study showing how to create a classification of software types using a curated set of software systems.

Our contributions show that existing, and very likely even new, classification attempts are deemed to fail for one or more issues, that we named as the `antipatterns' of software classification tasks. We collected 7 of these antipatterns that emerge from both our case study, and the existing classifications. 

These antipatterns represent recurring issues in a classification, so we discuss practical ways to help researchers avoid these pitfalls. It becomes clear that classification attempts must also face the daunting task of formulating a taxonomy of software types, with the objective of establishing a hierarchy of categories in a classification.

\end{abstract}


\begin{highlights}
\item Current software classification datasets lack generalizability.
\item We identified 7 antipatterns that pervade software classification. 
\item Hierarchical aggregation of categories can reduce some of the antipatterns.
\end{highlights}


\end{frontmatter}



\input{introduction}
\input{related}
\input{perils}
\input{case_study}
\input{conclusions}
 \bibliographystyle{elsarticle-num}
 \small
 \bibliography{cas-refs}





\end{document}

%% file: introduction.tex
\section{Introduction}

In the context of empirical software engineering research, the main goal of empirical papers is to achieve the generality of the results. The most common approach in doing so is to analyse projects having different application domains to decrease threats due to the generalizability of the results: as only a few examples, the work in~\cite{mojica2013large} analyses a collection of 200,000 mobile systems,~\cite{wen2019large} examples 1,500 GitHub systems based on their popularity, while~\cite{zhao2017impact} is based on 165,000 GitHub projects based on the Travis CI. As a side effect, the domain, context and uniqueness of a software system have not been considered very often by researchers as driving factors for detecting similarities or differences between software systems.

In parallel, there has been a call for `context-driven software engineering research'~\cite{briand2012embracing,briand2017case}: in the testing and verification fields, for example, the set of assumptions is obviously specific to the systems under test, and those assumptions are based on the type of system, the development process, and other factors. Although the diversity and context of software systems have received some attention in the past~\cite{vassallo2018context, easterbrook2008selecting}, contemporary research in the computing field is almost entirely application-independent. This has not always been the case - early in the computing era, `\textit{there were totally separate application domains (for example, scientific and data processing) and the research focus was often application-specific}'~\cite{glass1995contemporary}. 

 From the practitioners' point of view, categories and types of software systems are an important aspect to consider. Well known collaborative platforms like GitHub, that contain very large amounts of software repositories, show an increasing need to search and retrieve repositories based on their semantics. As a solution, GitHub has started to propose a service called \textit{Topics}\footnote{\href{https://github.com/topics}{https://github.com/topics}}, that allows developers to annotate their projects manually, and other users to search software via these topics. GitHub also provides the means to create \textit{Collections}\footnote{\href{https://github.com/collections}{https://github.com/collections}} (previously named Showcases), that is a curated list of topics where good quality repositories are grouped and showcased under the same umbrella. However, both these solutions have various issues: for example, \textit{Topics} are only optional features to a hosted software project, while GitHub does not suggest or restrict its usage in any way. As a result, there are plenty of similar (or identical, with a different morphological form) topics, making the search less effective. On the other hand, the \textit{Collections} list is manually curated; therefore, it is not scalable to all topics, reducing the effectiveness of finding repositories, especially those annotated with non-popular topics. \revision{Furthermore, developers tend to not use these tools, or using topics that are not helpful to retrieve their code (e.g., using programming languages).}

\revision{The call for context-driven software engineering research, easier retrieval of relevant projects using semantics, and the extra burden put on the developers to label their project with all the correct labels requires a more automated way to label software projects.} 

From past research and efforts, there have been several approaches to perform software classification, and depending on what seed classification has been used as a stepping stone. In some cases, the seed was initiated with a \textit{top-down} approach, i.e., using an external classification~\cite{vasquez2014api,soll2017classifyhub}: researchers would then use the categories (or labels) of the given classification to fit a sample of software projects. In other cases, categories were generated by the researchers~\cite{Borges2016popularity}, and the software projects assigned to the categories using again a top-down approach. Finally, a \textit{bottom-up} approach was used when researchers used, as categories, the labels assigned by developers to their own software~\cite{capiluppi2020towards}.

\revision{Moreover, there are various artefacts that can be used to perform the classification, from README files, to the source code. These two approaches are very different in terms of difficulty, for example the README might be lacking or containing irrelevant information about the repository content like information regarding the building of the code. Whereas the source code, there are hundreds or thousands of files, each containing some relevant semantic information that needs to be aggregated keeping track not only about the frequency but also about the interactions of the files.}

\revision{The misclassification, or lack of it, has various implications both on the repository, and also on the research that makes use of the labels. The developer might struggle to find contributors for their new and less popular repositories, as these are unable to discover and use the code. Furthermore, research that uses poorly labeled projects might infer wrong patterns and give bad advice to practitioners.}



In this work, we evaluate several existing software classifications proposed in the literature. The selection criteria of these works comprise \begin{enumerate*}[label=(\roman*)]
\item research papers that attempt a classification of application domains, and 
\item research works that made their data available.\end{enumerate*}
While analysing the resulting body of research works, we came across a number of recurring issues that researchers struggled with. These represent the most common \textit{antipatterns} in classifying software systems. 

Similarly to the work in \cite{kalliamvakou16promiseperils} that highlighted the pitfalls of mining software repositories, the goal of our work is to analyse existing classifications from past datasets, and to present a list of common antipatterns that researchers encountered when creating them. 

In this work we focus on the following research questions:

\begin{enumerate}[label=\textbf{RQ\arabic*} - , leftmargin=*]
\item \revision{How} is the \textit{quality} of existing software classification datasets?
\item \revision{What} are the \textit{antipatterns} of creating a software classification dataset or a taxonomy of software application domains?
\item How can we \textit{improve} software classifications and move towards a universal taxonomy that can be actively shared and used?
\end{enumerate}

\begin{figure}
    \centering
    \includegraphics[width=\columnwidth]{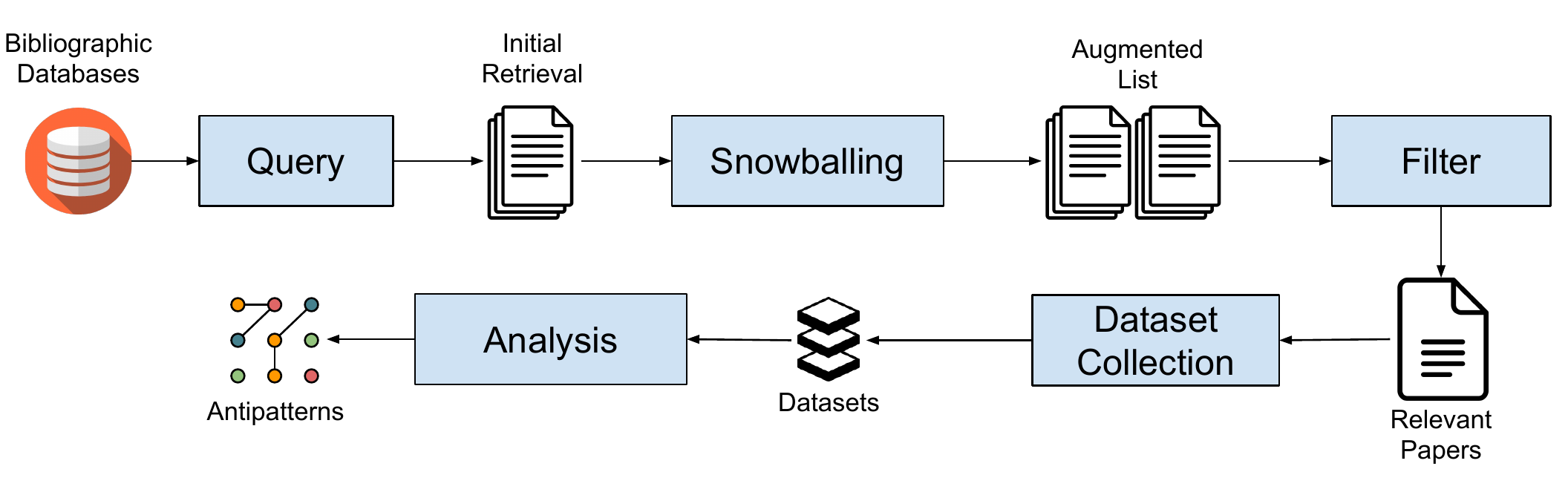}
    \caption{Pipeline used to identify the antipatterns in software application domain classification datasets.}
    \label{fig:pipeline}
\end{figure}

This paper presents two main contributions: first, we perform a case study attempting to create a classification for software systems that minimizes common issues present in current datasets' classification. Second, using the acquired experience and inductive analysis, we distil a set of 7 common \textit{antipatterns} that researchers have encountered while attempting to classify software systems. These antipatterns might have happened (and are likely to happen again) when the researchers \begin{enumerate*}[label=(\roman*)]
\item create their own labels, 
\item use a pre-defined classification of labels, or 
\item use the labels manually entered by software developers. 
\end{enumerate*} \revision{A visual representation of our pipeline is presented in Figure~\ref{fig:pipeline}.} For the sake of replication, we have made all our data\footnote{\href{http://doi.org/10.5281/zenodo.5018234}{http://doi.org/10.5281/zenodo.5018234}} and code~\footnote{\href{https://github.com/SasCezar/ComponentSemantics/blob/dev-semantic/componentSemantics/class\_term\_based\_classification.ipynb}{https://github.com/SasCezar/ComponentSemantics/blob/dev-semantic/componentSemantics/class\_term\_based\_classification.ipynb}} publicly available. 


The rest of this work is structured as follows: in Section~\ref{sec:background} we will give an overview of previous work with a focus on the used classifications. Using the evidence obtained from the analysis of the datasets, in Section~\ref{sec:pandp} we summarise the antipatterns  when creating a software classification. In Section~\ref{sec:dataset} we present a case study on creating a software classification and resolving some issues.  In Section~\ref{sec:threats} we discuss the threats to validity to our work. Finally, we present our conclusions and discuss future developments for our work in Section~\ref{sec:conclusions}.

%% file: related.tex
\section{Related Work and Existing Taxonomies}
\label{sec:background}

There have been several attempts in the literature focusing on software classification: in our paper we choose to only focus on those performing a classification of application domains. In general, all those previous works use their own datasets and different classifications. This generates an even more broad issue: it becomes hard to have a real applicability of these approaches, or an agreement on a shared benchmark.

While this paper is not a systematic literature review, the analyzed works have been selected using a similar approach to a systematic literature review. We retrieved the past works that (1) focused on the classification of software into application domains, and that (2) are 
\revision{proposing a new dataset.}
In order to perform this query, we performed an initial search \revision{on computer science bibliography services like \textit{dblp}, \textit{Google Scholar}, and \textit{Arxiv}. We used the} following terms: `\textit{software categorization}', `\textit{software classification}', `\textit{github repository classification}', and `\textit{software similarity}'.
We perform a first stage to validate the relevance of each work, we filter on the results using the title and abstracts. The works that passed the first filtering were subsequently used to perform a manual forward and backward snowballing for further relevant papers. \revision{Works that are case limits, were kept if their method or dataset can be used to perform software classification (e.g., \cite{theeten2019import2vec}, \cite{Borges2016popularity}).}

The list of papers that form the result of our search are listed in Table~\ref{tab:summary}, and it spans a window of 15 years (2006 to 2020). The approaches used to perform the classification task of software projects in the retrieved works varies: from project metrics~\cite{liu2018onboarding}, to source code~\cite{vasquez2014api}, and binary data~\cite{escobar2015bytecode}. In this paper, we focus on the approaches based on:

\begin{description}
    \item (A) source code; and
    \item (B) other project data (e.g., \textit{README files}),
\end{description}

as we are interested in the classification task using semantic information, and structural (can be extracted from source code). \revision{Table~\ref{tab:works} contains a list of the works divided by their approach.}

Below we provide, for each of the used information source a summary of the representative works with a focus on the work's dataset. 
A more in detail review of the software classification task landscape is presented in~\cite{auch2020similarityreview}.

\begin{table}[]
\centering
\caption{List of works divided by the different data source used.}
\label{tab:works}
\begin{tabular}{ll}
\toprule
Data Source                 & Works \\ \midrule
Source Code &  \cite{Kawaguchi2006MUDABlue, tian2009lact, vasquez2014api, leclair2018neural, mcmillan2012clan, linares2016clandroid, altarawy2018lascad, theeten2019import2vec, ohashi2019cnn_code}     \\
Other Project Data             &  \cite{vargas2015automatic, sharma2017cataloging, soll2017classifyhub, nguyen2018crosssim, zhang2019HiGitClass, sipio2020naive, izadi2020topic, Borges2016popularity}      \\ \bottomrule
\end{tabular}
\end{table}


\subsection{Source Code Approaches}
One of the initial works on software classification is MUDABlue~\cite{Kawaguchi2006MUDABlue}, which applied information retrieval techniques to classify software into 6 SourceForge categories. In particular, the authors used Latent Semantic Analysis (LSA), on the source code identifiers of 41 projects written in C. 

Following MUDABlue, Tian et al. proposed LACT~\cite{tian2009lact}, an approach based on Latent Dirichlet Allocation (LDA), a generative probabilistic model that retrieves topics from textual datasets, to perform the classification task from the identifiers and comments in the source code. In addition, the authors use a heuristic to cluster similar software. The authors use a dataset of 43 examples divided in 6 SourceForge categories. The list of projects is available in their paper. 

A different approach was adopted in~\cite{vasquez2014api}, the authors used API packages, classes, and methods names and extracted the words using the naming conventions. Using the example in~\cite{Ugurel2002classification}, the authors use information gain to select the best attributes as input to different machine learning methods for the task of classifying 3,286 Java projects into 22 SourceForge categories. Their dataset is not available anymore.

LeClair et al.~\cite{leclair2018neural} used a neural network approach. The authors use the project name, function name, and the function content as input to a C-LSTM~\cite{zhou2015clstm}, a combined model of convolutional and recurrent neural networks. Their dataset is made of 9,804 software projects, with the annotations from the Debian packages repository. The authors only analysed programs containing C/C++ source code, divided into 75 categories: many of these categories have only a few examples, and 19 are duplicate categories with different surface form, more specifically `contrib/X', where X is a category present in the list. 

CLAN~\cite{mcmillan2012clan} provides a way to detect similar apps based on the idea that similar apps share some semantic anchors. Given a set of applications, the authors create two terms-document matrices, one for the structural information using the package and API calls, the other for textual information using the class and API calls. Both matrices are reduced using LSA, then, the similarity across all applications is computed. 
Lastly, the authors combine the similarities from the packages and classes by summing the entries. The data is not available.
In \cite{linares2016clandroid}, the authors propose CLANdroid, a CLAN adaptation to the Android apps domain, and evaluate the solution on 14,450 Android apps. Their dataset is not available.

Another unsupervised approach was adopted by LASCAD~\cite{altarawy2018lascad}, a language agnostic classification and similarity tool. As in LACT, the authors used LDA over the source code, and further applied hierarchical clustering with cosine similarity on the output topic terms matrix of LDA to merge similar topics. The authors also proposed two datasets: an annotated one consisting of 103 projects divided in 6 categories (from GitHub Collections) with 16 programming languages (although many languages have only 1 example), and an unlabeled one which is not available. 

Taking a more Natural Language Processing (NLP) inspired approach, based on the distributional hypothesis: `\textit{A word is characterized by the company it keeps}'~\cite{firth1957studies}, \cite{theeten2019import2vec} proposed a neural network solution to create dense representation (i.e., embeddings) of libraries. The authors used the co-occurrences of import statements of libraries to learn a semantic space where libraries that appear in the same context are close (similar) in the space. The authors do not perform classification, therefore, their dataset is not annotated, however, the learned representation can be used to compute similarity and also train a classification model.

Differently from the previous works, \cite{ohashi2019cnn_code} used the C++ keywords and operators, represented as a binary matrix, as input to a convolutional neural network to assign the correct category out of 6 in the computer science and engineering field. The dataset is made of 40,023 students written source code for assignment/exams (short, single file, programs). Their dataset is not publicly available.

\subsection{Other Approaches}
The following is a review of the works that have been based on software artifacts other than source code. Following MUDABlue, Sally~\cite{vargas2015automatic} used an approach based on bytecode, the external dependencies of the project and information from Stack Overflow to generate a tag cloud. Their dataset is no longer available.  

Sharma et al.~\cite{sharma2017cataloging} used a combined solution of topic modeling and genetic algorithms called LDA-GA~\cite{panichella2013ldaga}. The authors apply LDA topic modeling on the README files, and optimize the hyper-parameters using genetic algorithms. While LDA is an unsupervised solution, humans are needed to annotate the topics from the identified keywords. The authors release a list of 10,000 examples annotated by their model into 22 categories, which was evaluated using 400 manually annotated projects. It is interesting to notice that half of the projects eventually end up in the `{Other}' category, which means that they are not helpful when training a new model.

ClassifyHub~\cite{soll2017classifyhub} used an ensemble of 8 na\"{i}ve classifiers, each using different features (e.g. file extensions, README, GitHub metadata and more) to perform the classification task. The authors use the InformatiCup 2017\footnote{\href{https://github.com/informatiCup/informatiCup2017}{https://github.com/informatiCup/informatiCup2017}} dataset, which contains 221 projects unevenly divided into 7 categories.

Nguyen et al.~\cite{nguyen2018crosssim} proposed CrossSim, an approach that uses the manifest file and the list of contributors of GitHub Java projects: this data is used to create a RDF graph where projects and developers are nodes, and edges represent the use of a project by another or that a developers is contributing to that project. The authors used SimRank~\cite{glen2002simrank} to identify similar nodes in the graph. According to SimRank, two objects are considered to be similar if they are referenced by similar objects. 

HiGitClass~\cite{zhang2019HiGitClass} used an approach for modeling the co-occurrence of multimodal signals in a repository (e.g. user, name of repository, tags, README and more). The authors performed the annotation according to a taxonomy (hierarchical classification) that is given as an input with keyword for each leaf node. The authors released a dataset with taxonomies for two domains: an artificial intelligence (AI) taxonomy with 1,600 examples, and a bioinformatics (Bio) one with 876 projects. 

Di Sipio et al.~\cite{sipio2020naive} used the content of the README files and source code, represented using TFIDF, as input to a probabilistic model called Multinomial Na\"{i}ve Bayesian Network to recommend possible topics. Given its premises, the work is defined as a multi-label classification. The authors used 120 popular topics from GitHub, and released a dataset of around 10,000 annotated projects in different programming languages.

Repologue~\cite{izadi2020topic} also adopted a multimodal approach. The authors used project names, descriptions, READMEs, wiki pages, and file names concatenated together as input to BERT~\cite{devlin-etal-2019-bert}, a neural language model, that creates a dense vector representation (i.e., embeddings) of the input text. Then,  a fully connected neural network was applied to these embeddings to predict multiple categories. Their dataset (currently unavailable) contains 152K repositories in various languages classified using 228 categories from GitHub's \textit{Collections}, which should be similar as the ones from Di Sipio et al.~\cite{sipio2020naive}.

Finally, Borges et al.~\cite{Borges2016popularity}, albeit not performing a classification of software repositories, made a list of 2,500 projects (annotated in 6 domains/categories) available for other researchers.

\subsection{Summary of Related Work} Table~\ref{tab:dataset} presents a summary of the datasets used in the literature. A single file with all the categories and the number of examples for each of the analyzed works is available in the replication package for inspection or further analysis. We use the following attributes to describe and analyze each dataset:
\begin{itemize}
    \item \textbf{Work}: the publication details  where the dataset is proposed or used for the first time;
    \item \textbf{Year}: the publication year;
    \item \textbf{Data Source}: the type of information used to perform the classification task from the software systems. This was further coded into the following:
    \begin{itemize}
        \item \textit{Source Code}: when the authors of the research directly used the source code of a system to infer (i.e., bottom up) or assign (i.e., top down) the software categories;
        \item \textit{README}: when the authors used the textual description of a software project of the README file to infer or assign one or more categories;
        \item \textit{Imports}: when the authors focused on what external libraries have been imported into a software project to infer or assign a category; 
        \item \textit{Key and Op}, when the authors used the predefined or reserved words of a specific programming language (e.g., C++), along with the operators used in the source code;
        \item  \textit{Multimodal}~\cite{Baltrusaitis2019multimodal}: when the authors used a combination of several sources (e.g., \textit{Source Code} and \textit{Wiki pages}).
 
    \end{itemize}
    
    \item \textbf{Available}: whether the dataset is available or not. We distinguish whether the available part is the list of annotated projects, or the list and the used files are;
    
    \item \textbf{Task}: the type of task that can be performed using the dataset. This attribute has the following possibilities:

    \begin{itemize}
        \item \textit{Classification}: assign one of $n$ mutually exclusive categories to the input project;
        \item \textit{Multi-Label Classification}~\cite{Tsoumakas2007MultiLabelCA}: assign to a project one or more categories from set of $n$;
        \item \textit{Hierarchical Classification}~\cite{gordon1987review}: assign $m$ of $n$ categories as for the Multi-Label problem, however there is a hierarchy among the categories;
        \item \textit{Similarity}: the task is to retrieve software that are similar to one given as input;
        \item \textit{Representation Learning}~\cite{bengio2013representation}: a more general case of the Similarity, in this case the goal is to create a dense representation (embedding) that preserves the similarities among projects, and it can also be used for downstream tasks.
    \end{itemize}

    \item \textbf{Examples}: the total amount of examples in the dataset;
    \item \textbf{Categories}: the number of different categories used to classify the software into, higher scores aren't always better as we will see later on;
    \item \textbf{Balance}: the level of class balance in terms of examples. It is computed using the Shannon Diversity Index (or Normalized Class Entropy~\cite{kalousis2004normalizedclassentropy} in Machine Learning), a normalized entropy~\cite{shannon1948mathematical} value:

\begin{equation*}
    \mbox{Balance} = \frac{-\sum\limits_{i = 1}^k \frac{c_i}{n} \log{ \frac{c_i}{n}}} {\log{k}}
\end{equation*}

where the numerator is the entropy for a dataset of size $n$ with $k$ categories each of size $c_i$, and the denominator is the perfect case of a dataset with balanced categories, used to normalizes the values. The results range between 0 (e.g., a completely unbalanced dataset with only one category containing all the examples) and 1 (e.g., a perfectly balanced dataset containing categories with the same amount of examples). A low score means that the dataset contains a large number of categories that are not well represented in examples, and therefore more difficult to perform the classification task when encountered. \revision{This measure is not suitable for cases where there is a large amount of classes with many examples, and only a few classes with a small number of examples;} 

\item \textbf{Min}: \revision{the number of examples for the class with the least amount of representation;}
\item \textbf{Max}: \revision{the number of examples for the largest class in the dataset.}




\end{itemize}

\newcolumntype{g}{l}

\begin{landscape}
\begin{table*}[htb!]
\footnotesize
\begin{center}
    \caption{Summary of the different datasets used in literature}
        \begin{tabularx}{1.65\textwidth}{llllllllgg}
        \toprule
        \multirow{2.5}{*}{Work} & \multirow{2.5}{*}{Year} & \multirow{2.5}{*}{Data Source} & {\multirow{2.5}{*}{Available}} & \multirow{2.5}{*}{Task} & \multicolumn{5}{c}{Dataset Stats} \\
        \cmidrule(lr){6-10}
                             &      &       &     &    & Examples       & Categories  & Balance  & Min & Max    \\ \midrule
MUDABlue~\cite{Kawaguchi2006MUDABlue} & 2006 & Source Code & List Only$^\diamond$ & Classification & 41 & 6 & 0.91 & 2 & 13 \\
LACT~\cite{tian2009lact} & 2009 & Source Code & Yes & Classification & 43 & 6 & 0.97 & 4 & 9 \\
CLAN~\cite{mcmillan2012clan} & 2012 & Source Code & No & Similarity & 8,310 & - & - & - & -  \\
Vasquez et al.~\cite{vasquez2014api} &  2014 & Source Code & No & Multi-Label Class. & 3,286 & 22 & 0.96$^{\dagger}$ & 303  & 1115 \\
Borges et al.~\cite{Borges2016popularity} & 2016 & - & List Only & Classification & 2500 & 6 & 0.88 & 103 & 837 \\
CLANdroid~\cite{linares2016clandroid} & 2016 & Multimodal & No & Similarity & 14,450 & - & - & - & -  \\
Sharma et al.~\cite{sharma2017cataloging} & 2017 & README & List Only & Classification & 10,000 (5,360$^\ddagger$) & 22 & 0.60 (0.91$^\ddagger$) & 85  & 670$^\ddagger$ \\
ClassifyHub~\cite{soll2017classifyhub}$^\divideontimes$  & 2017 & README & List Only &  Classification & 208 & 5 & 0.88 & 95 & 19 \\
LeClair et al.~\cite{leclair2018neural} & 2018 & Source Code & Yes & Classification & 9,804 & 75 & 0.73 & 1 & 3534 \\
LASCAD~\cite{altarawy2018lascad} & 2018 & Source Code & Yes & Classification & 103 & 6 & 0.95 & 7 & 26\\
CrossSim~\cite{nguyen2018crosssim} & 2018 & Multimodal & Yes & Similarity & 582 & - & - & - & - \\
Import2Vec~\cite{theeten2019import2vec} & 2019 & Imports & Embedding & Repr. Learning & - & - & - & - & - \\
Ohashi~\cite{ohashi2019cnn_code} & 2019 & Key and Op & No & Classification & 40,023 & 23 & 0.93 & 4713 & 10769 \\
{HiGitClass~\cite{zhang2019HiGitClass}} - AI & {2019} & {Multimodal} &  {Yes} & {Hierarchical Class.} & 1,596 & 3 - 13$^\bigstar$ & 0.58 - 0.87$^\bigstar$ & 48 - 1213$^\bigstar$ & 21 - 361$^\bigstar$ \\
{HiGitClass~\cite{zhang2019HiGitClass}} - Bio & {2019} & {Multimodal} &  {Yes} & {Hierarchical Class.} & 876 & 2 - 10$^\bigstar$ & 0.87 - 0.91$^\bigstar$ & 261 - 27$^\bigstar$ &  615 - 210$^\bigstar$ \\
Di Sipio et al.~\cite{sipio2020naive} & 2020 & README & Yes & Multi-Label Class. & 12,060 & 134 & 1 & 100 & 100 \\
Repologue~\cite{izadi2020topic} & 2020 & Multimodal & No & Multi-Label Class. & 152,000 & 228 & - & - & - \\ 
\textbf{Awesome-Java} & 2021 & - & Yes & Classification & 495 & 69 &  0.93 & 1 & 39 \\
\textbf{Reduced AJ} & 2021 & Source Code & Yes & Classification & 495 & 13 & 0.93 & 8 & 100\\
\midrule
\multicolumn{8}{l}{$\diamond$ A reproduced dataset is available in LACT~\cite{tian2009lact}} \\
\multicolumn{8}{l}{$\dagger$ Multi-labels, and also numbers in the paper table do not sum to the number of examples used for the measure}\\
\multicolumn{8}{l}{$\ddagger$ After removing the `Others' class that contains almost half of the examples in the dataset}\\
\multicolumn{8}{l}{$\divideontimes$ InformatiCup 2017 Dataset}\\
\multicolumn{8}{l}{$^\bigstar$ Two level hierarchy. First value is for the first level, the other for the second level of the taxonomy.}\\
\bottomrule
         \end{tabularx}
\label{tab:summary}
\end{center}
\end{table*}
\end{landscape}

\subsection{Related Work Content Analysis}
The quantitative summarization of the previous section is not sufficient to give us a complete idea of the datasets. In this section we present the content of the categorizations in the selected datasets. We will give an overview of their intended application, inferred from the labels, and discuss in more detail the semantics of the labels using word embeddings.

\begin{figure}[htb!]
    \centering
    \includegraphics[width=0.99\columnwidth]{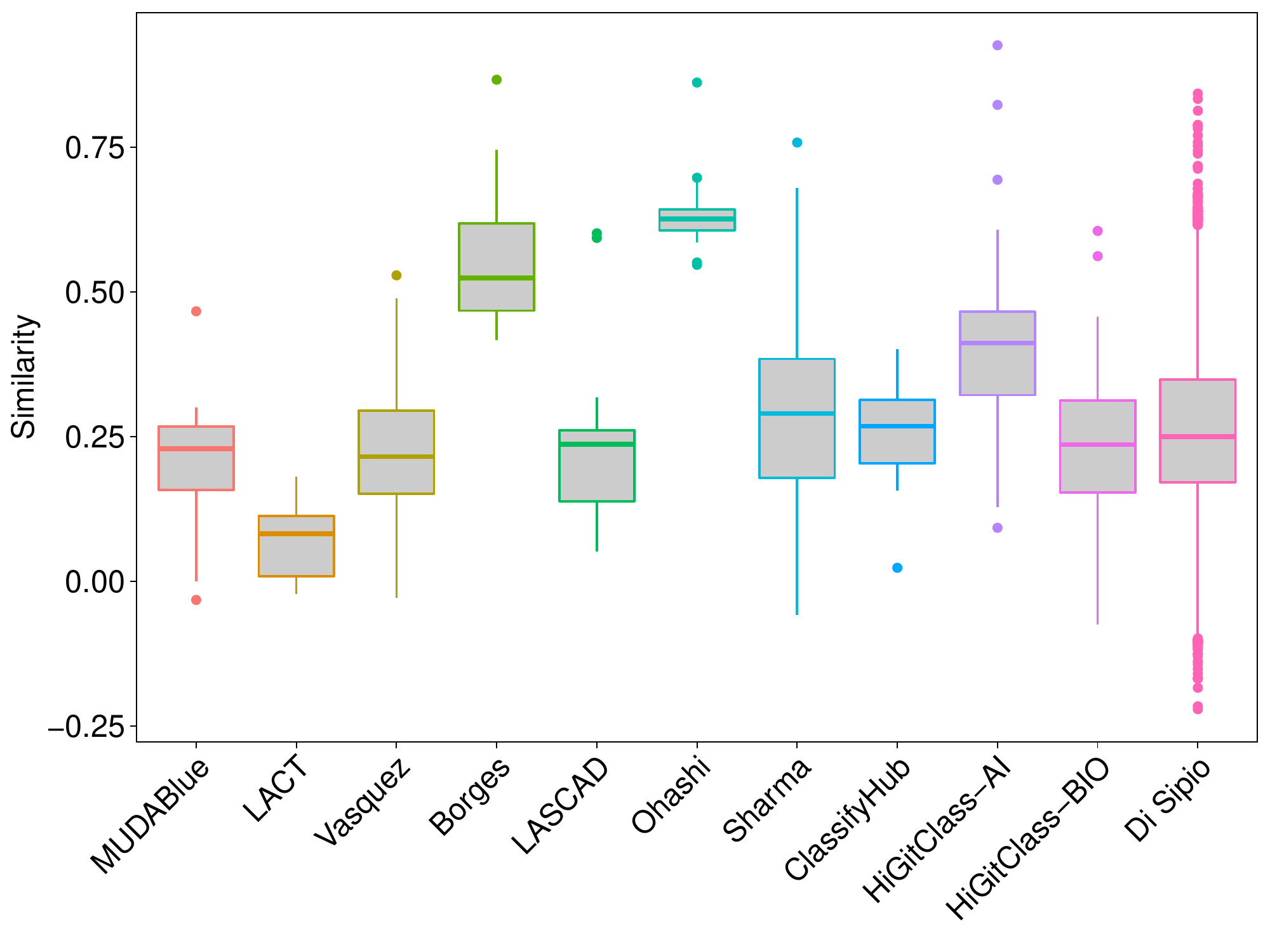}
    \caption{Cosine Similarity between labels using fastText embeddings.}
    \label{fig:label_similarity}
\end{figure}

We use fastText~\cite{bojanowski-etal-2017-enriching}, a neural language model, as the method for extracting the vector representation of the category: this is because it can handle out-of-vocabulary words, however, we obtained similar results also with BERT~\cite{devlin-etal-2019-bert} and StackOverflow~\cite{Efstathiou18SOW2V} embeddings. In Figure~\ref{fig:label_similarity}, we can see the distribution of similarities among categories, for each dataset. On the one hand, it is difficult to say anything definitive of the low similarity outliers, as terms from different domains might have low similarity; on the other hand, for the high similarity ones, these mostly result in categories that are in a hierarchical relationship or are highly related.


\begin{itemize}
    \item \textbf{MUDABlue}: it has a very small categorization, with a focus on developers containing categories like \dscat{Compilers}, \dscat{Editor}, and a very specific one \dscat{xterm}. However, it also contains labels that are not relevant to the others, in particular \dscat{Boardgame}. Overall the terms are not too similar between themselves, with the only outlier being the pair \dscat{xterm} and \dscat{Compilers} of $0.46$. And the lowest similarity being among the \dscat{Boardgame} label, and \dscat{Editor}. Given its small size, we can see the high spread as a lack of specificity.
    
    \item \textbf{LACT}: it has a similar domain as MUDABlue, but with more terms. It is also more general, as it contains several terms that are broader and less specific. We can find:  \dscat{Database} and \dscat{Editor}, as in MUDABlue, \dscat{Terminal}, which can be considered a more general version of \dscat{xterm}.  The other terms are more cohesive compared to MUDABlue, for example \dscat{E-Mail} and \dscat{Chat}. In this case, we do not find any outlier classes with a high similarity, and the distribution is quite narrow.
    
    \item \textbf{Vasquez}: it proposes a much more general taxonomy, as is a subset of SourceForge. Its labels span multiple fields in the Computer Science domain, some more general: \dscat{Scientific}, \dscat{Networking}, and \dscat{Security}: others more specific: \dscat{Indexing}, and \dscat{Compilers}. Given its well defined focus, and a higher number of topics compared to previous dataset, we find some labels that have a high similarity: \dscat{Compilers} and \dscat{Interpreters} result in a similarity of $0.52$ while \dscat{Networking} and \dscat{Communication} are at $0.48$. The latter are co-hyponyms, so hyponyms that share the same hypernym, while the former are related, as communication software use networking technologies.
    
    \item \textbf{LASCAD}: smaller compared to Vasquez et al., while still using Computer Science labels, they are not as complete and their labels are more sparse and less related to each other. The labels are: \dscat{Machine Learning}, \dscat{Data Visualization}, \dscat{Game Engine}, \dscat{Web Framework}, \dscat{Text Editor}, and \dscat{Web Game}. As expected from their surface form, the high similarity outliers pairs are: \dscat{Web Game} and \dscat{Game Engine}, with a similarity of $0.60$, and \dscat{Web Game} and \dscat{Web Framework} with $0.59$. These pairs, beside being related by sharing a term, are also related in terms of usage.
    
    \item \textbf{Ohashi}: this is a very specific categorization, based on the domain of science courses. The label set includes: \dscat{Combinatorial Optimization Problems}, \dscat{Number Theory Problems}, \dscat{Shortest Path Problems}. The overall high similarity between labels is due to the fact that all the labels contain the \dscat{Problems} term. 
    
    \item \textbf{Sharma}: it is a developers oriented classsification. The terms cover various areas, labels include: \dscat{Security}, \dscat{Music}, \dscat{Gaming and Chat Engines}, and \dscat{Blogging}. Furthermore, there also some programming languages like \dscat{Lua} and \dscat{Ruby related}. 
    
    \item \textbf{ClassifyHub}: as a more educational oriented dataset, its focus is not well defined, and it has high level labels in very loosely related domains: \dscat{Homework}, \dscat{Documents}, \dscat{Development}, \dscat{Education}, and \dscat{Website}. 
    
    \item \textbf{HiGitClass}: their two datasets are very specific, one focusing on AI subfields, while the other is focused on Bioinformatics. 
    
    Labels in the AI dataset include: \dscat{Computer Vision}, \dscat{NLP}, and \dscat{Speech} at level zero and \dscat{Image Generation}, \dscat{Super Resolution}, \dscat{Language Modeling} at the first level.  In Figure~\ref{fig:label_similarity}, we can see the similarity among the labels at all levels. The outliers are due to surface similarity among the labels (e.g., \dscat{Text Classification} and \dscat{Image Classification}). As expected, the average similarity is higher, given the very specific domain which as mentioned, also means that some words are present in multiple labels, increasing this score.
    
    In the Bioinformatics dataset, labels include: \dscat{Computational Biology} and \dscat{Data-Analytics} at level zero, and \dscat{Sequence Analysis}, \dscat{Database and Ontology}, and \dscat{System Biology} at level one. This dataset contains some labels that are representing two distinct concepts (e.g., \dscat{Database and Ontology}), these labels are less informative when used as we are unsure of which of the two concepts the annotated project belongs to. The outliers show similar characteristics as for the AI dataset.

    \item \textbf{Di Sipio}: their categorization is the most general as it is a subset of the most common GitHub Topics. 
    The topics include application domains like \dscat{Machine Learning}, \dscat{Database}, and \dscat{Operating System}. Moreover, we find programming languages like \dscat{Python} and \dscat{Java}, and also companies and services like \dscat{Google} and \dscat{AWS}. Given the large variety in labels, we also have many that are highly related to others. We start with \dscat{Cryptocurrency} and \dscat{Bitcoin} with similarity $0.84$; followed by a list of database-related labels with similarity in the range $0.75-0.80$, these are \dscat{PostgreSQL}, \dscat{SQL}, and \dscat{MySQL}, \dscat{NoSQL}, \dscat{MongoDB}. And also \dscat{Machine Learning} with \dscat{Deep Learning} having a similarity of $ 0.77$. 
    
\end{itemize}

A complete list of the statistics and labels for each dataset is available in our data replication package\footnote{\href{https://zenodo.org/record/5018234}{https://zenodo.org/record/5018234}}.

\subsection{Discussion}

We gathered several insights analysing the results collected in Table~\ref{tab:dataset}: first, approximately one in three of the datasets are not publicly available; similarly, the authors have only released the list of categories in one in three of the datasets, which in most cases is a sub-sample of a larger classification. In both those cases, it is hard (if not impossible) to reproduce the steps that lead to the classification: the unclear pre-processing has in fact a direct effect on the performance of the empirical approach~\cite{UYSAL2014104}. 

Second, we noticed the variance of the amount of examples and the resulting classifications: from 41 examples to 12K or 150K categories (although the latter are not publicly available), and from 6 to 134 or 228 (again, the latter are unavailable). The higher bound of these stats shows acceptable numbers  for both the number of example and the number of different categories. 

\revision{Furthermore, from the inspection of the categories shows some issues: in particular, they contain some categories that are not relevant to the intended use of the dataset: \textit{software application domain classification}.}


From the observations above, it becomes clear that most existing classifications have fundamental issues that prevent them from being further adopted by other researchers. While creating a new classification, one should not only be able to reproduce the steps performed by other researchers, but also annotate the aspects that might represent common antipatterns for others pursuing a similar goal. 

\revision{Next, in Section~\ref{sec:pandp}, we collect all the practical insights gained from analyzing the datasets and systematically present issues that we found in other classifications.}

%% file: perils.tex
\section{Antipatterns}
\label{sec:pandp}


Using the evidence observed during our case study, and an inductive analysis of the state of the art classification and taxonomies that comprises the information summarized in Table~\ref{tab:summary}, we highlight 7 antipatterns that researchers have faced so far while creating a taxonomy or a classification of software. We also add a discussion to each, and a suggested solutions to reduce the effect of each of these antipatterns.

The analysis was performed with a particular focus on the following characteristics:
\begin{itemize}
    \item Coverage: each classification has its own domain, that can be more specific or general. With this requirement, we can evaluate if there are missing categories given the domain of the classification, and therefore evaluate its completeness and usability in its domain; 
    \item Cohesion: this also relates to the domain of the classification, however, with this requirement, we try to assess if there are categories that do not belong to the domain;
    \item \revision{Consistency}: lastly, we check if the classification has any other issue \revision{that affect its consistency} like duplicate or overlapping categories, categories with a bad surface form (e.g. in \cite{leclair2018neural} there is `contrib/math'  and `math'), or any other abnormality. 
\end{itemize}

The results of the analysis is a list of 7 common antipatterns. Below, we present a definition of each, and we will discuss them with instances of how each was observed in past literature and give some suggestions on how they can be fixed from a practical point of view.



\begin{description}
    \item[MT --] \textbf{Mixed Taxonomies}: this issue happens when a label set contains various categories from different taxonomies (e.g., Programming Languages and Application Domains).
    
    \item[MG --]\textbf{Mixed Granularity}: this issue emerges when categories are found (or imposed) by researchers to a dataset, although those categories belong to different levels (e.g., family, species, etc.) of a taxonomy.
    
    \item[SC --] \textbf{Single Category}: this is a very common issue in software classifications, and it is based on simplifying a complex artifact like a software system, and describing it with only one category.
    
    \item[NE --] \textbf{Non Exhaustive Categories}: this issue is visible when the categories chosen (i.e., top-down) or extracted (i.e., bottom-up) by the researchers don't cover the entire spectrum of software categories.
    
    \item[NRC --]\textbf{Non Relevant Categories}: this issue is visible when researchers choose or extract a subset of categories that is not representative of the domain of the classification.
    
    \item[UJC --]\textbf{Unnecessarily Joined Categories}: this issue occurs when researchers arbitrarily join or use several categories as one, although those are a compound of two or more different domains. 

    \item[SKC --] \textbf{Sink Category}: this is another very common issue in software classification, and it manifests itself when researchers use a generic category for all the software systems that do not fit any of the available categories.
\end{description}

\subsection{Mixed Taxonomies - MT}

\begin{description}[leftmargin=0cm]
\item\textit{Definition:} The MT antipattern is defined as a label set consisting of a mixture of two or more different taxonomies, each covering a different domain. 


\item\textit{Possible Root Cause:} We hypnotize, that one of the causes of this, is the fear of the creators of the dataset to exclude labels, which might make it less appealing to the final users. 
 
\item\textit{Potential Side Effects:}
This is problematic as the model has to perform two or more tasks at the same time with only a single label per project. Having multiple annotations for the same project is not a problem itself, the issue is having them mutually exclusive as there might be labels diverging from each other.

\item\textit{Concrete Examples:}
One very common additional part of the classification is based on having programming languages in the labels set. This is common to~\cite{sharma2017cataloging, leclair2018neural, sipio2020naive}, where, on average, $10\%$ of the examples belong to programming language categories. It is also common to have language specific frameworks or technologies as part of the label set, as for example found in \cite{sipio2020naive}. In some cases, we might find a domain category like `Deep-Learning', and specific framework or technology labels like `Django' or `AWS' that are part of the same classifications.

\item\textit{Potential Solution:}
A solution to this issue is to define the different classifications independently, and use them as separate tasks when training models. Having projects annotated with different classifications is useful as they can be used as auxiliary tasks to a multi-task learning~\cite{caruana1997mtl} model to improve generalizations~\cite{ruder2017overview}, and boost performance for all the tasks.
\end{description}

\subsection{Mixed Granularity - MG}

\begin{description}[leftmargin=0cm]
\item\textit{Definition:} 
Having a dataset where some labels are very specific for a field and others are more general, or worse, when labels are in an `\texttt{IS-A}' relationship, without these relations being explicitly represented. 

\item\textit{Possible Root Cause:} This issue is caused by the difficulty in creating such relations among all the labels in the categorization.

\item\textit{Potential Side Effects:} 
 The former can make the model catch very specific terms, that are dependant on the sample in the dataset, to distinguish between categories. The latter causes overlap between classes, making the classification harder or even impossible when having a single annotation for the projects.

\item\textit{Concrete Examples:}
As examples of this antipattern in action, we observed that \cite{sipio2020naive} contains the `Cryptocurrency' and `Bitcoin' categories, that have a similarity of $0.84$. Similarly, \cite{vasquez2014api} contains the `Compilers' and `Interpreters' categories with a similarity of $0.52$. Even in the label sets where we could not detect hierarchical relations among categories (for example in~\cite{vasquez2014api}), we also observed more general categories like `Networking' and `Web' along with very specific ones like `Interpreters' and `Indexing'.

\item\textit{Potential Solution:} A solution to this antipattern is to perform a refinement of the categories, and try to aggregate them, in a hierarchical fashion, as we attempted in our case study. The benefits are visible in Figure~\ref{fig:label_similarity}, we have a lower number of outliers with high similarity. Moreover, a more qualitative analysis to evaluate the extent of the issue, is to assign a position in a taxonomic ranking scale, like the ones used in biology, for each category and evaluate the number of different ranks covered by the used taxonomy. 
\end{description}

\subsection{Single Category - SC}

Software systems do not contain only one large feature or functionality, but they are rather composed of many other smaller parts, each with its own specific task. Most of the times software systems are labeled with only one category, that limits the extent to which researchers can learn from it. 

\begin{description}[leftmargin=0cm]
\item\textit{Definition:} We define the \textit{Single Category} antipattern as the annotation of software with only one category, despite it being a mix of different ones.

\item\textit{Possible Root Cause:} This is caused by the annotation process being time consuming and also by not being made by the developers, who will only need to annotate their own project.

\item\textit{Potential Side Effects:}
Different components of a project, each with its own domain, influence what a model learns, making it harder for it to assign a single category to all components, especially when their semantic contribution to the model is different. This antipattern gets more evident when having a Mixed Granularity (and also a Mixed Taxonomy) classification, where one category is contained in another, however the system is penalized by suggesting the other category.

\item\textit{Concrete Examples:}
This antipattern was detected in all datasets except for the one of Vasquez et al.~\cite{vasquez2014api} that performs multi-label classification, and \cite{sipio2020naive, izadi2020topic} that performs recommendation of GitHub \textit{Topics}. 

\item\textit{Potential Solution:}
 While the solution for this is obvious (e.g., `annotate each sample with multiple categories'), this is not that easy to achieve as it requires extra effort from researchers and developers during the annotation phase. A less demanding approach to achieve the goal would be to adopt the annotation approach as in the GitHub \textit{Topics}, however those topics are highly noisy as mentioned previously. Therefore, this antipattern requires more attention in future works.
\end{description}

\subsection{Non Exhaustive Categories - NE}
\begin{description}[leftmargin=0cm]
\item\textit{Definition:} A taxonomy where there are terms that have a common parent, but one of them is lacking, is considered to be suffering of this anti-pattern. For a classification to be usable, it needs to cover the entire range in the domain. This is dependant on the actual domain, and also changes over time, and therefore has to be considered in the domain behind the classification. 

\item\textit{Possible Root Cause:}
There are various possible causes, one is that the missing term was not existent, or very uncommon, at the time that the taxonomy was created (e.g., the Deep Learning was not very common 20 years ago). Another cause is the how the taxonomy was created, if a subsampling of another one, then some terms might have been excluded by the process, if the taxonomy is defined top-down, then the knowledge of the domain by the authors has a big impact on the presence or not of this antipattern. 

\item\textit{Potential Side Effects:}
Having a classification that is too small or with missing relevant categories is an issue as the classification model performance can be affected based on what category is missing, as the category can be easily differentiated if no similar ones are present. Moreover, will make the approach less useful for general uses. 

\item\textit{Concrete Examples:}
Some example from the previous work are the classifications of \cite{Kawaguchi2006MUDABlue, tian2009lact, altarawy2018lascad} that are too small and lack many categories, and in \cite{leclair2018neural}, where they have `Interpreters' but not `Compilers', they are also missing a `Security'/`Cryptography' related categories. In~\cite{sipio2020naive}, a GitHub Topics based classification, we have `NLP' but not `Computer Vision' in spite of it having 134 categories.

\item\textit{Potential Solution:}
 Solutions to this are limited, as mentioned previously, application domains change over time with new categories appearing. A possible solution is to have a coarser granularity, however, this might not be a possibility in some cases, and will also reduce the utility of the classification.

\end{description}

\subsection{Non Relevant Categories - NRC}



\begin{description}[leftmargin=0cm]
\item\textit{Definition:}
 This antipattern is based on assigning very fine-grained categories to a project. This means that researchers have in the past added categories to their taxonomies that are too specific and non relevant. 

\item\textit{Possible Root Cause:} The presence of these categories can be caused by a lack of a specific usage for the taxonomy; or a lack of refinement of categories, when subsampling them from a larger pool.

\item\textit{Potential Side Effects:}
This antipattern has the effect of making the classification task too simple, since the categories have very few shared terms (or even none) with the others. This can be viewed as a special case of Mixed Taxonomies: however, the categories that would be more relevant to those are usually one or two, and, differently from Mixed Taxonomies, they are not related to specific technologies or programming languages.

\item\textit{Concrete Examples:}
 Examples of non relevant categories were found for instance in \cite{Kawaguchi2006MUDABlue}, where categories are very different between each other (e.g., `Boardgame' and `Editor`); and in \cite{soll2017classifyhub}, where there are a `Development' category and a `Homework' one. Another example is from \cite{sipio2020naive}, where we find `Minecraft' (i.e., a popular videogame) as a category.

\item\textit{Potential Solution:}
 A possible workout for this antipattern would be to just remove these categories and either discard the examples along with it, or try to reassign them to a relevant one that belongs to the domain that is being modeled.
\end{description}

\subsection{Unnecessarily Joined Categories - UJC}



\begin{description}[leftmargin=0cm]
\item\textit{Definition:}
This antipattern manifests itself in categories that join several categories using the ``and'' conjunction (e.g., `Gaming and Chat Engines`). While this is a less common antipattern, having a category that is a conjunction of two unrelated categories is something to pay attention to.   

\item\textit{Possible Root Cause:}
One cause for this is the high similarity between the terms, and the low number of examples each of them have, therefore joining them will make the number of examples higher.

\item\textit{Potential Side Effects:}
The joint categories do not provide as much of an information to the final user as having a single term: which category of the two in the conjunction does the project belong to?.

\item\textit{Concrete Examples:}
In~\cite{sharma2017cataloging} there are many examples of these joined categories: while some might be considered acceptable (for example, `{Data Management and Analysis}') others form a weak combination (for example, `Gaming and Chat Engines` or `Build and Productivity tools'), since they join labels that belong to very different categories. 

\item\textit{Potential Solution:}
An easy solution for this antipattern would be for researchers to avoid using conjunctions, or to use them only when the categories are related. However, if the categories are indeed related, there should be a more general (single) label to group them under, which is a more appropriate solution.

\end{description}

\subsection{Sink Category - SKC}

\paragraph{Definition:}
This is a very common antipattern to fall in, when dealing with large classifications. This antipattern manifests itself with a category, used as a super-label, that is applied to any software that doesn't fit any another category in the classification, but that still needs an annotation. The most common one is to have a category named `Others', or other synonyms. However, there are other categories that might not be that obvious, like `Frameworks', `Libs' and so on. 

\paragraph{Possible Root Cause:} While the `Other' category is needed for the classification, it's abuse and presence of the oter \textit{Sink Categories} is eased by the difficulty of annotating some projects, hence labeling them with a \textit{Sink Category} makes it easier.

\paragraph{Potential Side Effects:}
This sink category adds extra noise, as it might get applied to the majority of projects contained in the pre-existing classification. This category might be also applied to projects that actually belong to other categories, but that were not originally contained in the classification; and also can be used as a backup for harder to classify projects.

\paragraph{Concrete Examples:}
Examples from previous work include: in LeClair et al.~\cite{leclair2018neural} has three of these which are `Libs', `Utils', and `Misc' which total to $30\%$ of the dataset size; in \cite{Borges2016popularity} they have a category called `Non-web libraries and Frameworks' containing $25\% $ of their dataset's examples. Lastly, \cite{sharma2017cataloging} has a category `Others' containing  50\% of the dataset examples.   

\paragraph{Potential Solution:}
 This antipattern is a harder to avoid, and it was commonly found in our survey: the works that do not suffer from this were usually dealing with small classifications, or very domain-specific ones.  

\color{black}

\subsection{Summary}
In Table~\ref{tab:pathologies} we summarize, for each work, the antipatterns they have in their classification.  We can notice the least easy to fall in antipatterns are the NRC and UJC, and the most common are NE and SC which are also the hardest to avoid as they require extra work in the annotation phase. The most problematic issues, MT and MG are also quite common, with the former being present in most of the larger and more general taxonomies.

We can also see that there is no perfect taxonomy: if one only considered the amount of antipatterns contained in a dataset, they would select the works of Di Sipio et al.~\cite{sipio2020naive}, Zhang et al. (HiGitClass)~\cite{zhang2019HiGitClass}, and Ohashi et al.~\cite{ohashi2019cnn_code}. However, the latter two have very specific and closed domains, that are more straightforward to create, but less useful to other researchers.

\begin{table}[htb!]
\centering
    \caption{Summary of the antipatterns in previous works.}
    \begin{tabular}{lccccccc}
&  \rot{MT} & \rot{MG} & \rot{SC} & \rot{NE} & \rot{NRC} & \rot{UJC} & \rot{SKC} 
        \\
        \midrule
MUDABlue~\cite{Kawaguchi2006MUDABlue}     & \OK & \OK & \OK  & \OK  & \OK  &      &      \\
LACT~\cite{tian2009lact}                  &     & \OK & \OK  & \OK  &      &      &      \\
Vasquez et al.~\cite{vasquez2014api}      &     & \OK &      &      &      &      & \OK  \\
Borges et al.~\cite{Borges2016popularity} &     & \OK & \OK  & \OK  &      &      & \OK  \\
LeClair et al.~\cite{leclair2018neural}   & \OK & \OK & \OK  &      &      &      & \OK  \\
LASCAD~\cite{altarawy2018lascad}          &     & \OK & \OK  & \OK  &      &      &      \\
Ohashi et al.~\cite{ohashi2019cnn_code}   &     &     & \OK  & \OK  &      &      &      \\
Sharma et al.~\cite{sharma2017cataloging} & \OK &     & \OK  & \OK  &      & \OK  & \OK  \\
ClassifyHub~\cite{soll2017classifyhub}    & \OK &     & \OK  & \OK  & \OK  &      &      \\
HiGitClass~\cite{zhang2019HiGitClass}     & \OK &     & \OK  &      &      & \OK     &      \\
Di Sipio et al.~\cite{sipio2020naive}     & \OK & \OK &      & \OK  &      &      &      \\
        \bottomrule
    \end{tabular}
\label{tab:pathologies}
\end{table}

\revision{In Section~\ref{sec:dataset}, we present an attempt at creating a classification using a new, bottom up taxonomy: we will annotate all the steps in doing so, and try and address the limitations of existing classifications presented above.}

%% file: case_study.tex
\section{Case Study}
\label{sec:dataset}





In this section we describe our attempt at creating an \revision{example} classification, \revision{with real world usages}, that minimises the general issues noted above. 
This section describes the original source of the classification (\ref{subsec:class_source}), the manual process that was used to reduce the categories in order to balance the number of examples in each (\ref{subsec:label_mapping}). Finally, in order to evaluate how distinct the categories are from each other, we evaluated the lexical similarity between categories, described by their projects content (\ref{subsec:lexsimil}).

\subsection{Classification Source}
\label{subsec:class_source}
As a starting point (e.g., the `seed') for the creation of our \revision{case study} dataset, we picked a pre-existing classification, from a \textit{Java Awesome List}  hosted on GitHub. Awesome Lists are curated repositories containing resources that are useful for a particular domain: in our case we use \textit{Awesome-Java}\footnote{\url{https://github.com/akullpp/awesome-java}}, a GitHub project that aggregates overall 700 curated Java frameworks, libraries and software organised in 69 categories. In an initial phase of cleaning, we removed tutorials and URLs to websites, obtaining 530 examples; we also removed the projects that could not be analyzed \revision{(e.g., gives errors in the pipeline, including: encoding, no keywords left to create an encoding, etc.)}. The total of projects finally considered for our task was 495. 

Using GitHub Topics\footnote{\url{https://github.com/topics}} could be an alternative to the selected Java Awesome List: however, the categories for the same list of projects is larger in the former (around 1,000 labels) than in the latter (69 labels). Also, the decision of using the Awesome Java list was to avoid using pre-existing classifications or taxonomies. Beside the previously mentioned issues, other have sporadically emerged in the past (e.g., in \cite{leclair2018neural}, where many examples of that dataset come from secondary code that is not relevant to the main projects. Moreover, \textit{Awesome-Java} is an annotation of a closed ecosystem (Java development), making it the seed of a small, but realistic, classification. In fact this process, when improved and automated, could be applied to GitHub's \textit{Topics} annotations to obtain an unlimited source of distantly annotated examples.  \revision{Lastly, the \textit{Awesome-Java} repository is a collective effort of more than 300 contributors and continuous updates to the list, making it the go-to source for more than 31K (as stars) developers when looking for a library.}

\revision{However, the GitHub Topics, would be a better source for a more general list of categories and a larger scale source of projects. However, there are larger challenges as there are more than 100K GitHub Topics, hence, this will move the focus of our study to unrelated issues.}

\subsection{Label Mapping}
\label{subsec:label_mapping}
The \textit{Awesome-Java} classification contains 69 categories: on average, each category contains 8 projects. Also, some of the categories represent either  general concepts (`Science') or  detailed keywords (e.g., `Bean Mapping'). 

As a result, the Awesome-Java categories make classification tasks quite challenging: therefore we decided to manually reduce the original categories, in order to reduce the complexity, and avoid duplicates or synonyms. This mapping was performed manually, in a hierarchical fashion, by one of the authors, and resulted in a smaller set of 13 categories (\textit{Reduced AJ}): the \textit{Label} column of Table \ref{tab:dataset} lists the reduced categories that were obtained from the original 69. \revision{The reductions were evaluated by the second author, and disagreements on terms were resolved by discussion.}

\begin{table}[htb!]
\small
\begin{center}
        \caption{Distribution of the number of examples for each category in the \textit{Reduced AJ}.}
        \begin{tabular}{lcc}
        \toprule
\textbf{Label}                  & \textbf{Projects} & \\
\midrule
Introspection          & 32    &    \\
CLI                    & 8     &    \\
Data                   & 49    &    \\
Development            & 100   &    \\
Graphical              & 11    &    \\
Miscellaneous          & 59    &    \\
Networking             & 25    &    \\
Parser                 & 41    &    \\
STEM                   & 39    &    \\
Security               & 14    &    \\
Server                 & 37    &    \\
Testing                & 42    &    \\
Web                    & 38    &    \\
\midrule
\textbf{Total} & \textbf{495} \\
\bottomrule
        \end{tabular}
\label{tab:dataset}
\end{center}
\end{table}

\begin{figure}[htb!]
    \centering
    \includegraphics[width=0.7\columnwidth]{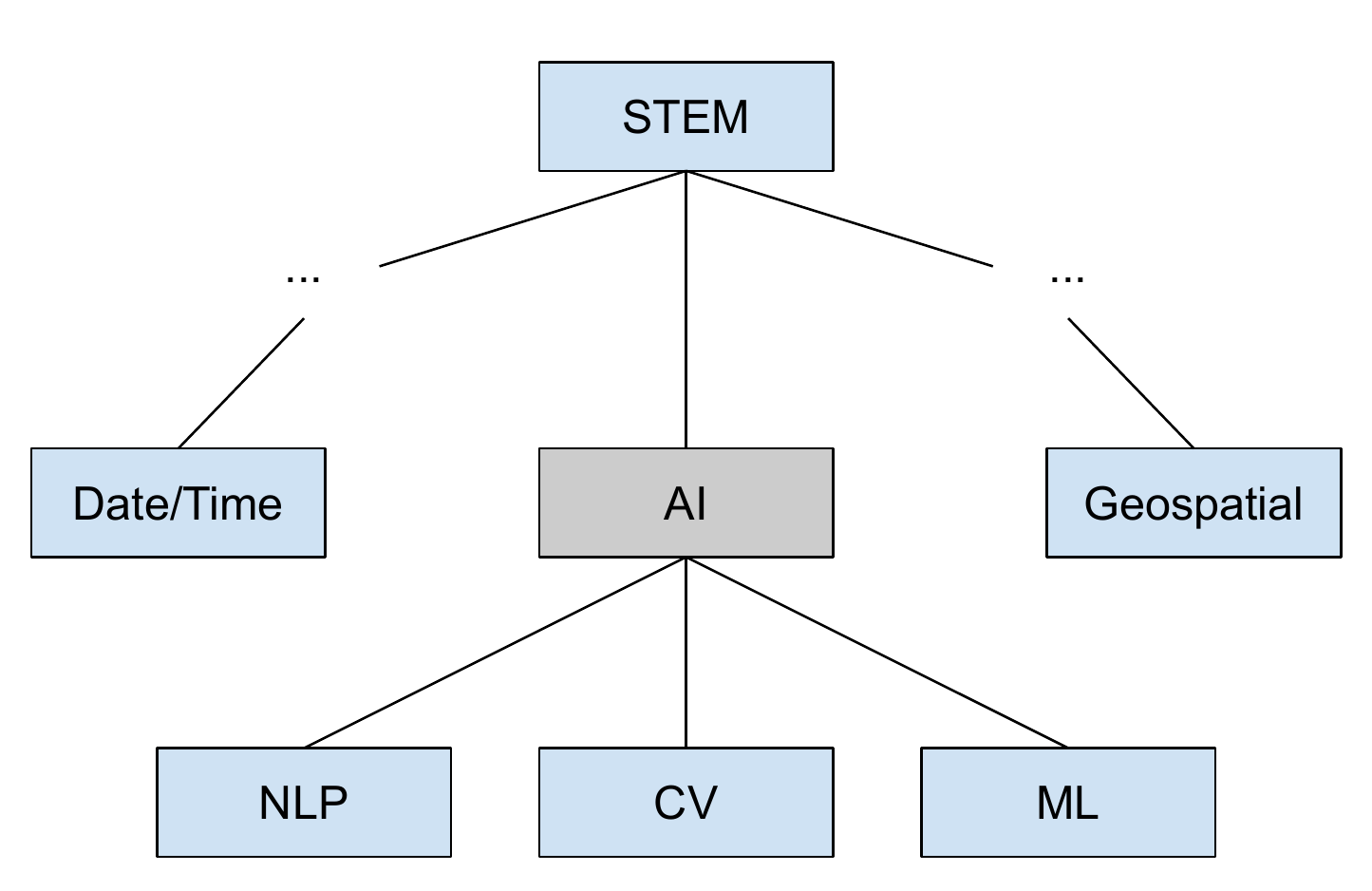}
    \caption{Example of the reduction process. The blue rectangles are actual classes in the initial or final dataset, the grey one are intermediate logical classes used to aggregate labels. The dots represent not well defined intermediate classes.}
    \label{fig:reduction}
\end{figure}

This reduction, in addition to increasing the amount of examples per class, also helps with one of the issues that the original \textit{Awesome-Java} presented, that is the lack of a hierarchical relation between categories. 
Figure~\ref{fig:reduction} shows a visual representation of the reduction, and how this helps the establishment of hierarchical links. Given three labels in the \textit{Awesome-Java} taxonomy: Natural Language Processing (`NLP'), Computer Vision (`CV'), and Machine Learning (`ML'); we reduce those three to an intermediate conceptual label `AI'. Together with two other categories, `Date/Time' and `Geospatial', we assign all these to the `STEM' category. 

The initial and final annotated labels are stored as a CSV file, and is available in our replication package. The file has the following schema:
\begin{itemize}
    \item \textbf{project.name}: name of the project;
    \item \textbf{project.desc}: short description of the project from \textit{Awesome-Java};
    \item \textbf{project.link}: URL to the GitHub repository;
    \item \textbf{category}: \textit{Awesome-Java} annotation;
    \item \textbf{category.desc}: short description of the category from \textit{Awesome-Java};
    \item \textbf{label}: mapping of the original category into one of the reduced set.
\end{itemize}

\subsection{Evaluation}
\label{sec:evaluation}

We evaluate the quality of our approach of reduction on the \textit{Awesome-Java} taxonomy using both qualitatively and quantitative measure. We first compare the original and the reduced taxonomies using the introduced perils and pitfalls. Furthermore, we measure the lexical similarity of classes.

\subsubsection{Antipatterns}

A summary of the antipatterns found in the original \textit{Awesome-Java} and the \textit{Reduced AJ} is present in Table~\ref{tab:pathologies_our}. The original \textit{Awesome-Java} presents several of the antipatterns identified in the taxonomies in previous works, from non exhaustive label set (NE), to mixed granularity (MG) and mixed taxonomies (MT). Therefore, we can be more confident that the processes we used to reduce these issues can be deployed to different taxonomies as well. Examples of the antipatterns found in \textit{Awesome-Java} include:

\begin{itemize}[noitemsep]
    \item Mixed Taxonomy: examples of this antipatterns are the presence of technologies like \dscat{Apache Commons} in the list;
    \item Mixed Granularity: for examples we find the label \dscat{Science} with the label \dscat{Configuration}, or \dscat{Development} and \dscat{Compiler Compiler}. Moreover, there are labels that are in a `\texttt{IS-A}' relationship, like \dscat{Mobile development} and \dscat{Development}.  
    \item Non Exhaustive Categories: one examples is the lack of a \dscat{Audio Processing} category, while there are for \dscat{Computer Vision} and \dscat{Natural Language Processing}. 
    \item Sink category: the \dscat{Apache Commons} label contains many projects that can be annotated with another label in the set, for example \dscat{Commons CLI\footnote{\href{https://commons.apache.org/proper/commons-cli/}{https://commons.apache.org/proper/commons-cli/}}} can be annotated with the \dscat{Command Line Interface} label.
\end{itemize}

The two main benefits of the reduction process are the removal of the Non Exhaustive (NE) label set issue and the removal of the Mixed Taxonomy (MT). Another benefit, although not completely removed as seen in Table~\ref{tab:pathologies_our}, is a marked decrease in the severity of the Mixed Granularity (MG) issue.

In our case study, most of these antipatterns have been resolved using the label reduction process. However better results require tackling Mixed Taxonomy (MT) issue: its resolution required manual annotations of the examples belonging to the problematic category (e.g., `Apache Commons') as the mere reduction would just map everything to Sink Category.

\begin{table}[htb!]
\centering
    \caption{Summary of the antipatterns in the original \textit{Awesome-Java} and our Reduced AJ.}
    \begin{tabular}{lccccccc}
&  \rot{MT} & \rot{MG} & \rot{SC} & \rot{NE} & \rot{NRC} & \rot{UJC} & \rot{SKC} 
        \\
        \midrule
\textbf{Awesome-Java}  &  \OK   & \OK  & \OK  & \OK   &      &    & \OK \\
\textbf{Reduced AJ}    &     & \OK  & \OK  &    &      &    &     \OK \\
        \bottomrule
    \end{tabular}
\label{tab:pathologies_our}
\end{table}

\begin{figure}[htb!]
    \centering
    \includegraphics[width=\columnwidth]{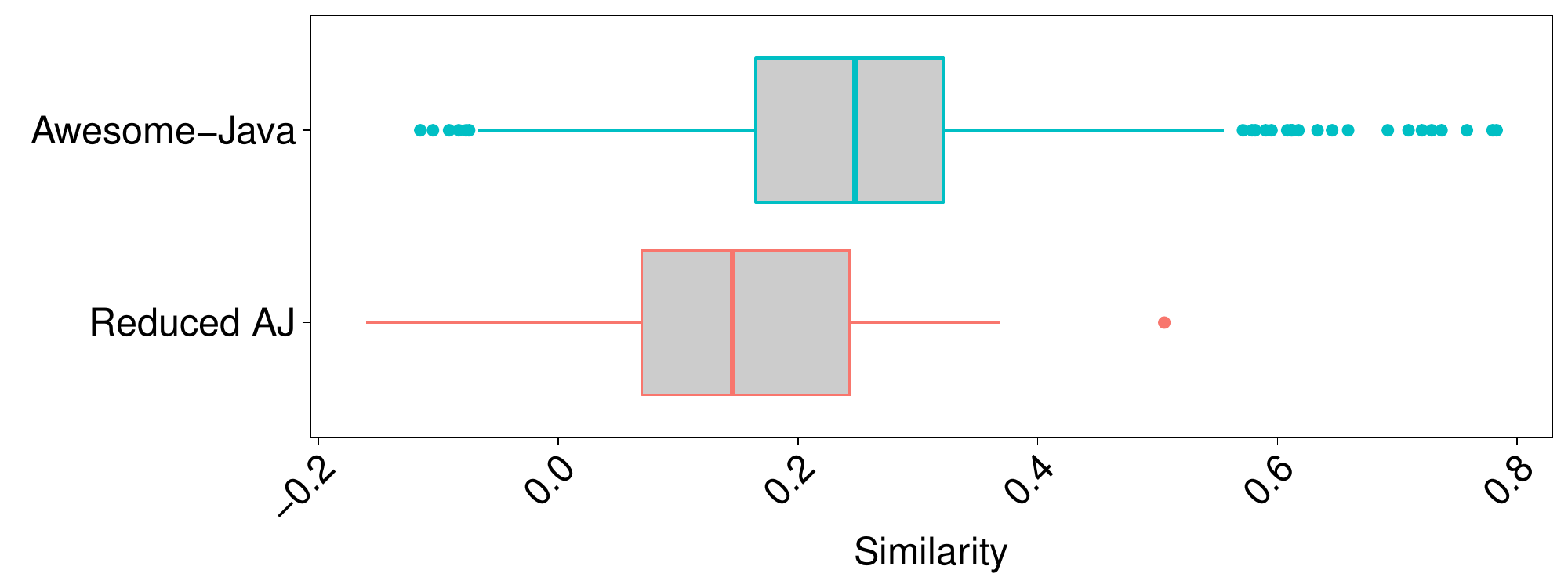}
    \caption{Cosine Similarity between labels using fastText embeddings.}
    \label{fig:label_similarity_our}
\end{figure}

Similarly for the other datasets, we also computed the similarity of the labels using fastText, in \ref{fig:label_similarity_our} we can see the similarity before and after the reduction. The lower average similarity is caused by a reduction in the terms in a hierarchical relationship, and also by a lower number of terms sharing a common subword.

\color{black}
\subsubsection{Lexical Similarity between Categories}
\label{subsec:lexsimil}

To evaluate the quality of our process, we evaluated the lexical similarity between each category using the content of all projects belonging to that category. This step is of fundamental importance, since it helps to evaluate the quality of the mapping process into categories, and to give an empirical evaluation of how similar categories are, before and after reduction. In order to lexically represent the categories we used the TFIDF approach; in order to measure the similarity of two categories, we used the cosine similarity. \revision{We did not opted for embedding solutions like fastText or BERT as they are not suited for our task. For example fastText is not designed for long document embeddings, as it performs a mean operation over the embeddings to create the final representation, meaning that all the documents will converge to a very similar embedding, resulting in very high similarities between all documents. With BERT, given the small amount of token it accepts as input (512) we will have a similar issue, as we will need to combine the embeddings of subset.  Other, more code oriented solutions, like code2vec or CodeBERT have issue as well. Code2vec is trained with the objective of encoding structure and semantic of the code, and not semantic of the word. CodeBERT suffers of the issues of BERT.}

\paragraph{Extraction of the category documents} For each project belonging to a category, we created the category document using all the \textit{identifiers} contained in the source code files of the project belonging to that category. For the extraction of the identifiers, we used the \textit{tree-sitter}\footnote{\href{https://github.com/tree-sitter/tree-sitter}{https://github.com/tree-sitter/tree-sitter}} parser generator tool. The identifiers, without keywords, are extracted from the annotated concrete syntax tree created using a grammar for Java code. 

The identifiers were further processed by (1) separating the camel case strings into words, (2) lower casing every word, and (3) removing common Java terms that do not add much semantically (e.g., `\textit{main}', `\textit{println}', etc). Lastly, we perform (4) lemming, which is a way to reduce amount of different terms in the vocabulary by removing the morphological differences in words with the same root (e.g., `\textit{networking}' becomes `\textit{network}').

\paragraph{Evaluation of the similarity between categories} These category documents were used as an input to TFIDF, a statistical vectorization of words based on the Bag of Words (BoW) model. Documents are considered as a collection of words/terms, and converted to a vector by counting the occurrences of every term. Differently from BoW, in TFIDF the words are weighted by their total frequency in the collection of documents. This will result in a list of vectors representing the lexical content of that category. We limit the amount to the top 1,000 terms that have a max document frequency lower than 0.8, hereby words that are present in less than 80\% of the labels, therefore ignoring common words.

We adopted the cosine similarity, a measure of similarity between two vectors, in order to measure the similarity between all categories, and to evaluate possible overlaps or large differences between them. The cosine similarity, when using TFIDF, ranges from 0, different content, to 1 identical content. We compute the similarities between the categories of the original finer grained \textit{Awesome-Java} classification, and the \textit{Reduced AJ} as well. 

\begin{figure*}
    \centering
    \includegraphics[width=\columnwidth]{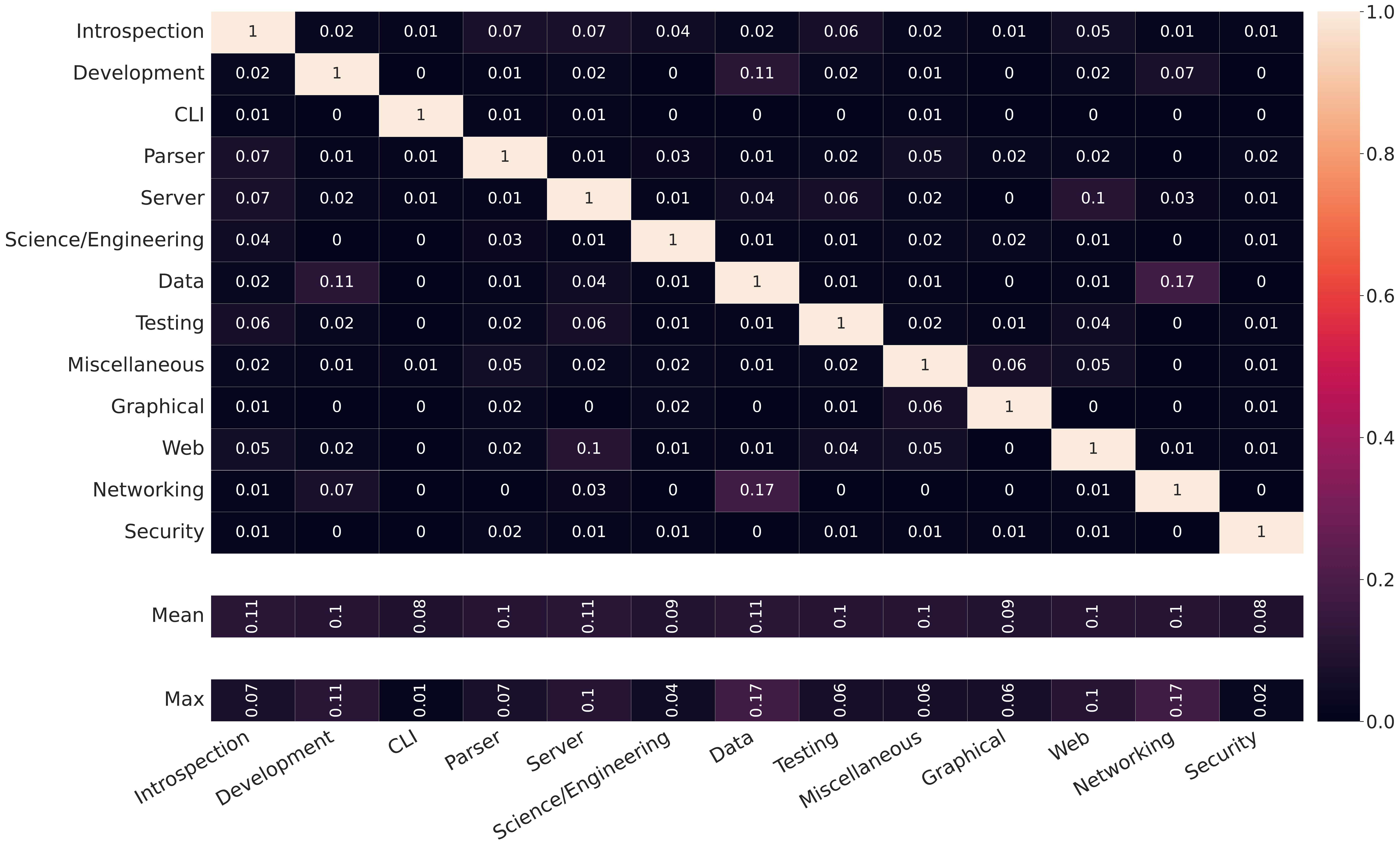}
    \caption{Cosine similarities between categories in the \textit{Reduced AJ}. The last two rows are the mean and max similarity per category.}
    \label{fig:class_sim}
\end{figure*}

\begin{sidewaysfigure}[!htp]
    \includegraphics[width=1.15\textwidth]{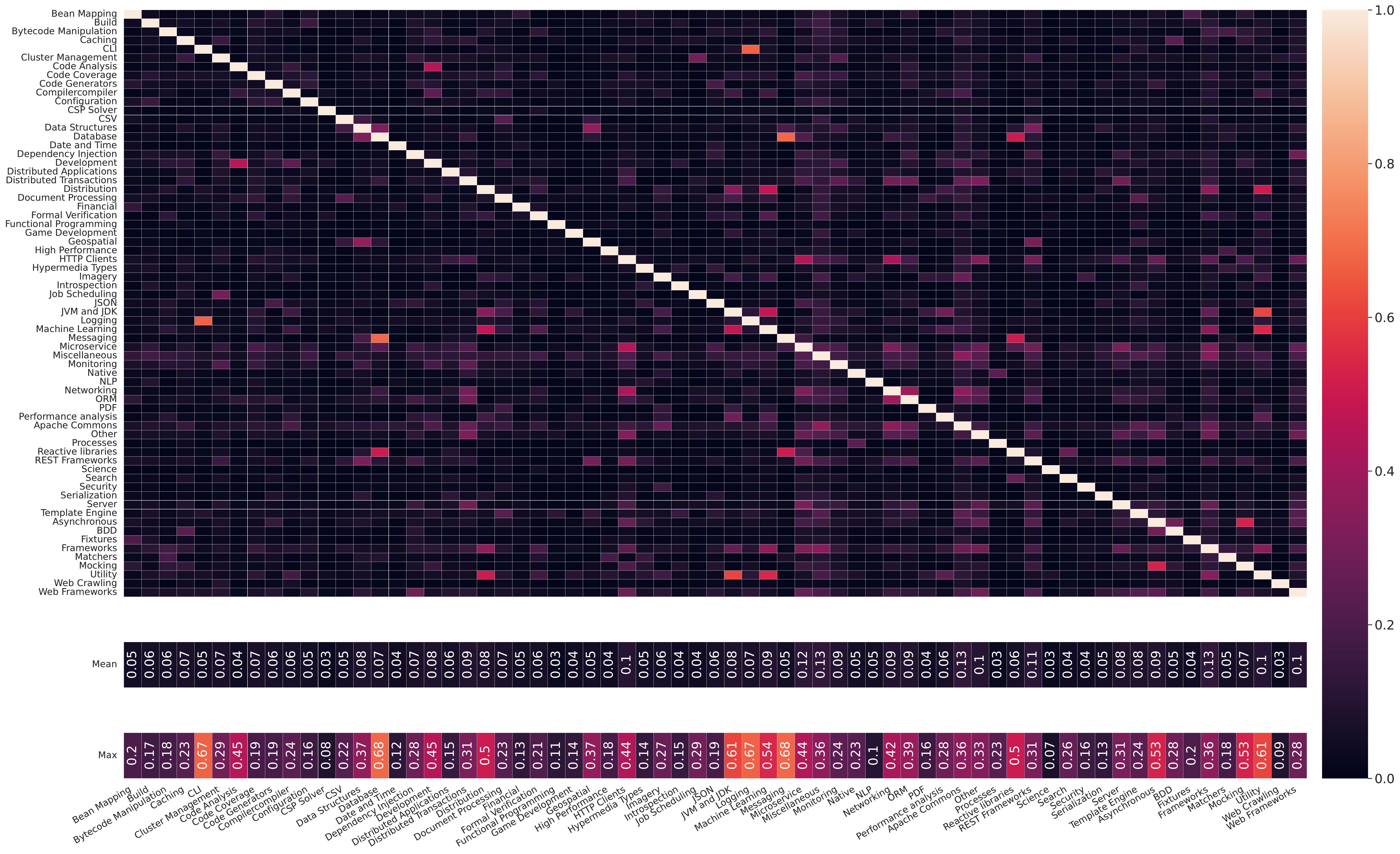}
    \caption{Cosine similarities between categories in the original \textit{Awesome-Java}. The last two rows are the mean and max similarity per category.} 
    \label{fig:class_sim3}
\end{sidewaysfigure}

\paragraph{Results} The results in Figure~\ref{fig:class_sim} show the final similarities for the reduced classification, while Figure~\ref{fig:class_sim3} shows the similarities between the categories of the original \textit{Awesome-Java}. The initial impression is that the overall similarity between categories is very low for both classifications: this is a clear effect of the pre-filtering of the terms that are very frequent in all documents. 
The second observation is that the mean similarity of the classification with 69 labels is higher and has more variance: in particular there is an average similarity of 
$0.0520\pm0.0677$, as compared to the reduced one of $0.0210\pm0.0290$. This is also visible graphically in the heat map of Figure \ref{fig:class_sim3}: the brighter spots, therefore higher similarity are much more frequent there than in the reduced classification in Figure~\ref{fig:class_sim}. 

\paragraph{Discussion} The higher similarities of Figure~\ref{fig:class_sim3} are caused by a combination of different factors: 
\begin{enumerate}
    \item the first cause is the presence \revision{of the Mixed Granularity antipattern}t. For example, the similarity between the categories `Development' and `Code Analysis' is 0.45. These two were mapped into the `Development' combined category of the reduced classification. 
    
    \item the second cause of high similarity \revision{is the Single-Label antipattern}. As a result, some projects are labeled with one category, but their features would require multiple labels. An example of this would be the high similarity (0.68) between `Database' and `Messaging' which in \textit{Awesome-Java} is described as ``\textit{Tools that help send messages between clients to ensure protocol independency}". An example explaining this high similarity is by considering `\textit{Apache Kafka}', a distributed event streaming framework used for various tasks including data integration, being categorized as `Messaging' while still containing a high amount of data management terms like `\textit{query}'. This also remains in the reduced classification (Figure~\ref{fig:class_sim3}) for the `Database' and `Networking', in which `Messaging' is mapped. 
    
    \item lastly, we also have to consider noise, given the smaller number of examples per category in the original classification, the documents used might not be very representative of the category.
\end{enumerate}

\subsection{Discussion: moving towards a complete taxonomy}
The purpose of a classification, like the ones that we have summarised in Table~\ref{tab:summary} is to organise similar items (in this case, software systems) into categories for future reference. This could be for proactively recommending systems~\cite{nguyen2018crosssim}, in order to generate a list of alternatives from the same category; or for identification of common patterns into the found categories~\cite{1d3929b4506d48aa949bf6a3ecf16d39}.


On the other hand, the purpose of a taxonomy is to organise categories into levels: for instance, in a taxonomy for biology,  `class' (say, ``mammals") is placed on a different level than `order' (say, ``carnivorous"). In the classification works that we analysed, we never observed an attempt to define at which level the categories are (except in ~\cite{zhang2019HiGitClass}), or whether those should be considered more or less generic or specific in the more general terms of a taxonomy.

As a further analysis, and in order to evaluate how specific, general, or mixed level the taxonomy is, we asked a group of 10 people \revision{belonging to our research group. The pool of annotators is} composed of PhDs, Post Docs, and Professors in the Software Engineering field to indicate whether the categories illustrated in Table~\ref{tab:dataset} should be placed in a higher or lower level of a taxonomy. The questionnaire included \revision{13 questions, one for each topic in the \textit{Reduced AJ}, where the annotators were asked to perform a rating by assigning the topic into one of} 5 levels, from 1 (very generic) to 5 (very specific). We collected their responses and analysed them to determine if any of the categories that were reduced from the Awesome-Java sample should be considered a `family', or `group' or even a `species' within a software taxonomy. 

\begin{figure}[!htb]
    \centering
    \includegraphics[width=0.7\columnwidth]{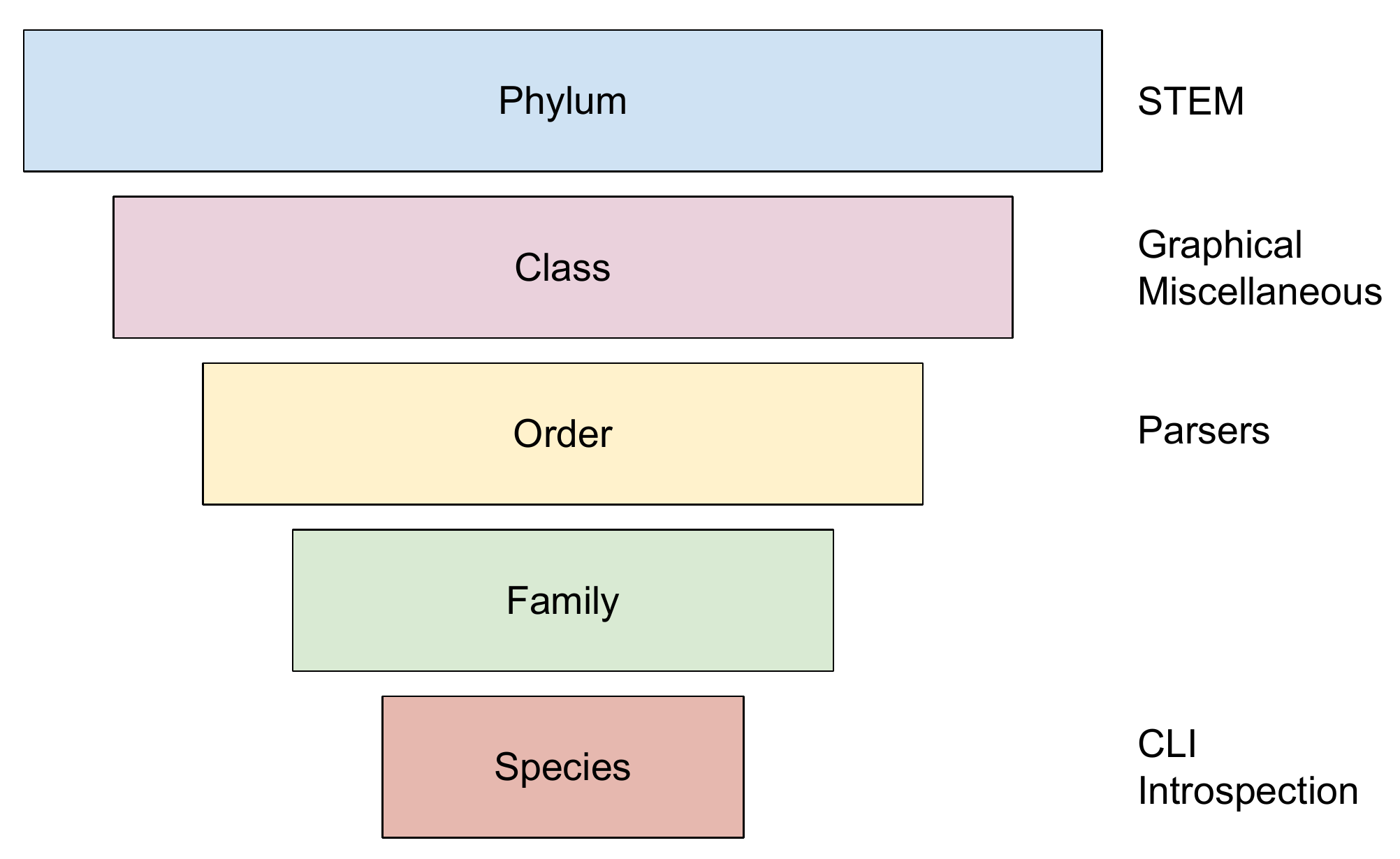}
    \caption{Assignment of the categories to levels of a taxonomy}
    \label{fig:taxo}
\end{figure}

The results of this preliminary qualitative analysis showed that specific categories were placed fairly consistently among either the very general (e.g., the `STEM' category), or the very specific level (`Introspection', `CLI'). On the other hand, several other categories were assigned uniformly to levels 2, 3 and 4, therefore being placed to middle-ground levels of the taxonomy, depending on the assessor's point of view. Figure~\ref{fig:taxo} shows the visualisation of what has initially emerged from the answers of our questionnaire. \revision{The assignment of a topic to a level was performed using the majority voting, those without majority are not presented.}

This is further evidence that defining categories for software systems faces the challenging task of placing them in an overarching map: the `mixed levels' antipattern will always affect a classification effort, unless a more concerted research effort is conducted and shared, in order to build a taxonomy and to place the categories in its levels. \revision{A magnifier of the the uniform distribution of some topics can be imputed to our methodology, the rating task is more complex compared to a ranking to asses subjective characteristics~\cite{ye2014subjective}. Hence, the future work will focus on using better methods to rank topics.}

%% file: conclusions.tex
\section{Threats to Validity}
\label{sec:threats}
We use the classification of Runeson et al.~\cite{Runeson2012threats} for analyzing the threats to validity in our work.  We will present the \emph{construct validity}, \emph{external validity}, and \emph{reliability}. Internal validity was not considered as we did not examine causal relations \cite{Runeson2012threats}.

\subsection{Construct Validity}
A construct threat in our work is the choice of the classification for the case study. However, given the wide variety of datasets, and the similarity Awesome Java has regarding the issues with the state of the art classifications, this threat is mitigated. 

Another threat regards the way the reduction was performed. Having a single annotator performing the reduction can increase the bias in the selection of the resulting categories. We mitigated this threat by having another author evaluate the resulting categories; furthermore, we collected feedback from other colleagues regarding the same resulting categories.
 
\subsection{External Validity}
We reduce the external validity to a minimum by analyzing a large variety of datasets. We analyzed 12 datasets, with a different origin of the base classification: both bottom up, and top down classifications have been considered for study. Moreover, these classifications are based on different domains, some more specific (e.g., bio-engineering) other more generic: this should help alleviate this threat to validity. 

\subsection{Reliability}
The analysis of the classifications and taxonomies is inherently subjective, as it involves natural language, and prior knowledge about the different application domains. We adopted objective tools, like semantic analysis, to aid with the subjective analysis.

\section{Conclusions and Future Work}
\label{sec:conclusions}

In this work we evaluated the different classifications used for the software classification task. The current classifications have issues that might compromise generalizability of classification models, moreover, there is no general classification that can be actively used (\textbf{RQ1}). We identified a list of 7 common antipatterns that researchers encounter when creating a software classifications for classifying systems into application domains (\textbf{RQ2}). While the ideal case would be to avoid those antipatterns when creating a classification, this is quite difficult, and a refinement stage helps with the reduction (but not the complete removal) of some of these issues. We presented a case study, using a real classification, in which 
we mitigated some of the antipatterns using a reduction of the categories (\textbf{RQ3}). The reduction was performed manually in a hierarchical fashion.

As future works we plan to perform the similarity between the categories content also for the other works in the literature. Furthermore, we plan to 
perform analysis similar to the work of Sen et al.~\cite{priyanka2020what} for the Question Answering (QA) task in the Natural Language Processing field, where they look at what clues QA models actually use to answer questions. We are interested in checking if the models learn general terms of a specific domains, or they pick up dataset specific clues that are not transferable to others.

We are also planning to create a taxonomy induced from all the GitHub \textit{Topics} given the variety of projects, and therefore application domains, that are hosted on the platform. 
Given the large amount of terms, the hierarchical aggregation process needs to be automated. First, we plan to create a ranking with a larger pool of annotators, and given the high disagreement in our ranking case study, use a different methodology: ranking from pairwise comparisons, as is less complex for annotators~\cite{shah2016estimation}. Lastly, use the ranking to create links between levels, to group terms from different levels in the same domain.